**COVID Future Panel Survey: A Unique Public Dataset Documenting How U.S. Residents' Travel-Related Choices Changed During the COVID-19 Pandemic**

**Rishabh Singh Chauhan**
PhD Candidate
Department of Civil, Materials, and Environmental Engineering
University of Illinois at Chicago
Chicago, IL 60607
Email: rchauh6@uic.edu

**Matthew Wigginton Bhagat-Conway**
Assistant Professor
Department of City and Regional Planning
University of North Carolina
Chapel Hill, NC 27599
Email: mwbc@unc.edu

**Tassio Magassy**
PhD Candidate
School of Sustainable Engineering and the Built Environment
Arizona State University
Tempe, AZ 85281
Email: tmagassy@asu.edu

**Nicole Corcoran**
PhD Student
School of Geographical Sciences and Urban Planning
Arizona State University
Tempe, AZ 85281
Email: ncorcor1@asu.edu

**Ehsan Rahimi**
Postdoctoral Research Fellow
Department of Civil, Materials, and Environmental Engineering
University of Illinois at Chicago
Chicago, IL 60607
Email: erahim4@uic.edu

**Abbie Dirks**
Masters Student
School of Sustainable Engineering and the Built Environment
Arizona State University
Tempe, AZ 85281
Email: acdirks@asu.edu

**Ram M. Pendyala**
Professor
School of Sustainable Engineering and the Built Environment
Arizona State University
Tempe, AZ 85281
Email: Ram.Pendyala@asu.edu

**Abolfazl (Kouros) Mohammadian**
Professor
Department of Civil, Materials, and Environmental Engineering
University of Illinois at Chicago
Chicago, IL 60607
Email: kouros@uic.edu

**Sybil Derrible**
Associate Professor
Department of Civil, Materials, and Environmental Engineering
University of Illinois at Chicago
Chicago, IL 60607
Email: derrible@uic.edu

**Deborah Salon**
Associate Professor
School of Geographical Sciences and Urban Planning
Arizona State University
Tempe, AZ 85281
Email: dsalon@asu.edu

Word Count: 6455 words + 4 table (250 words per table) = 7,455 words

**ABSTRACT**
The COVID-19 pandemic is an unprecedented global crisis that has impacted virtually everyone. We conducted a nationwide online longitudinal survey in the United States to collect information about the shifts in travel-related behavior and attitudes before, during, and after the pandemic. The survey asked questions about commuting, long distance travel, working from home, online learning, online shopping, pandemic experiences, attitudes, and demographic information. The survey has been deployed to the same respondents thrice to observe how the responses to the pandemic have evolved over time. The first wave of the survey was conducted from April 2020 to June 2021, the second wave from November 2020 to August 2021, and the third wave from October 2021 to November 2021. In total, 9,265 responses were collected in the first wave; of these, 2,877 respondents returned for the second wave and 2,728 for the third wave. Survey data are publicly available. This unique dataset can aid policy makers in making decisions in areas including transport, workforce development, and more. This article demonstrates the framework for conducting this online longitudinal survey. It details the step-by-step procedure involved in conducting the survey and in curating the data to make it representative of the national trends.

**Keywords:** Longitudinal Survey, COVID-19, Travel Behavior, United States.

## INTRODUCTION AND BACKGROUND

The COVID-19 pandemic has affected virtually everyone on the planet. The first confirmed case of COVID-19 in the United States (US) was reported in January 2020 (1). Over the next 30 months, more than a million people died from COVID-19 in the United States alone (2). It has impacted every aspect of life, from environment to economy, from mental health issues to societal aspects, and from travel to globalization.

Understanding and recording people's perceptions and changes in lifestyle and behaviors is crucial to ensuring a smooth transition to new habits and for policymakers to properly plan for what is to come. Aiming at investigating these patterns, surveys have been conducted all over the world exploring the pandemic's effects on various aspects of life, such as health, finance, and transportation. The pandemic has been volatile and unpredictable in terms of new outbreaks, variants, and restrictions. Therefore, peoples' opinions also shifted over time, and thus time-sensitive snapshots from a survey could be considered outdated in a matter of months. A great way to overcome this issue is by investigating how behavior changes over time through a panel survey, where the same respondents participate in multiple waves of a survey conducted at different time periods. Given its complexity and cost, this methodology is not frequently seen in the literature; however, such panel survey may provide unique information about the dynamics of peoples' behaviors and opinions changing over time during uncertain times such as the COVID-19 pandemic.

A few COVID-19 related panel surveys have been conducted in the US and around the globe. While some surveys have a general focus (3), others focus specifically on transportation (4). Drummond and Hasnine (5) studied the changes in shopping behavior in New York City in 2020. They found that increased subway usage was correlated to in-store shopping and that lower-income individuals were more likely to shop in-store. Results from another longitudinal survey in 2019 and 2020 showed that although online shopping increased during the early stages of the pandemic, long-term online shopping impacts should be modest (6). Another panel study conducted in Turkey during the early stages of the pandemic showed a sizeable decline in public transport usage and an important shift to working from home (7). Some changes in travel behavior are expected (e.g., decline in transit usage due to the increased risk of infection). Understanding other trends such as changes in work-from-home and increase in vehicle distance traveled of specific demographic groups would require further investigation. As these relationships cannot be seen simplistically, associating these changes to other characteristics such as demographic attributes, current mandates in places, attitudinal and psychological effects, and activity risk perceptions are meaningful information that can be associated to these societal changes. These relationships help explain and predict how people may respond to potential social disturbances they may encounter in the future.

Given how all these new habits are related to one another, having a comprehensive survey instrument that allows many crosstabulations and exploration of a number of potential relationships between variables of different nature is crucial to fully understand the changes that have happened, as well as to predict how people may behave in the future. However, most surveys do not allow such analyses as they tend to be single-themed. A more multidisciplinary survey instrument that is more robust and richer is arguably preferable but less common. Although surveys are great tools to understand public opinion and preference, most surveys found in the literature present some key limitations such as being specific in one single topic; having a localized sample, which prevents national representativeness; small sample size; or not covering key time periods.

Filling this gap in the literature to collect time-sensitive, detailed, and multidisciplinary data, we conducted the 'COVID Future Survey' – an online US-wide longitudinal survey that comprises questions on a wide range of topics. The survey was taken by the same respondents at different times and allows the study of behavioral and attitudinal shifts over time. The data were collected in three waves: the first wave was conducted from April 2020 to June 2021, the second wave from November 2020 to August 2021, and the third wave from October 2021 to November 2021. Figure 1 shows the timeline of the three survey waves along with several major pandemic related events. In total, 9,265 responses were collected in the first wave. Only these respondents were invited to participate in waves 2 and 3. A total of 2,877 respondents returned for the second wave and 2,728 for the third wave. In all, 1,933 respondents participated in all three survey waves.

The survey consists of questions related to commuting, long distance travel, working from home, online learning, online shopping, pandemic experiences, attitudes, and demographic information. The survey asked questions focusing on three timeframes: (a) before COVID-19, (b) during COVID-19, and (c) expectations for a time when COVID-19 will no longer be a threat. The questions in the first wave focus on all three timeframes. The second and third waves, however, ask questions only about during COVID-19 and post-COVID-19 since the information about pre-COVID-19 was already collected in the first wave.

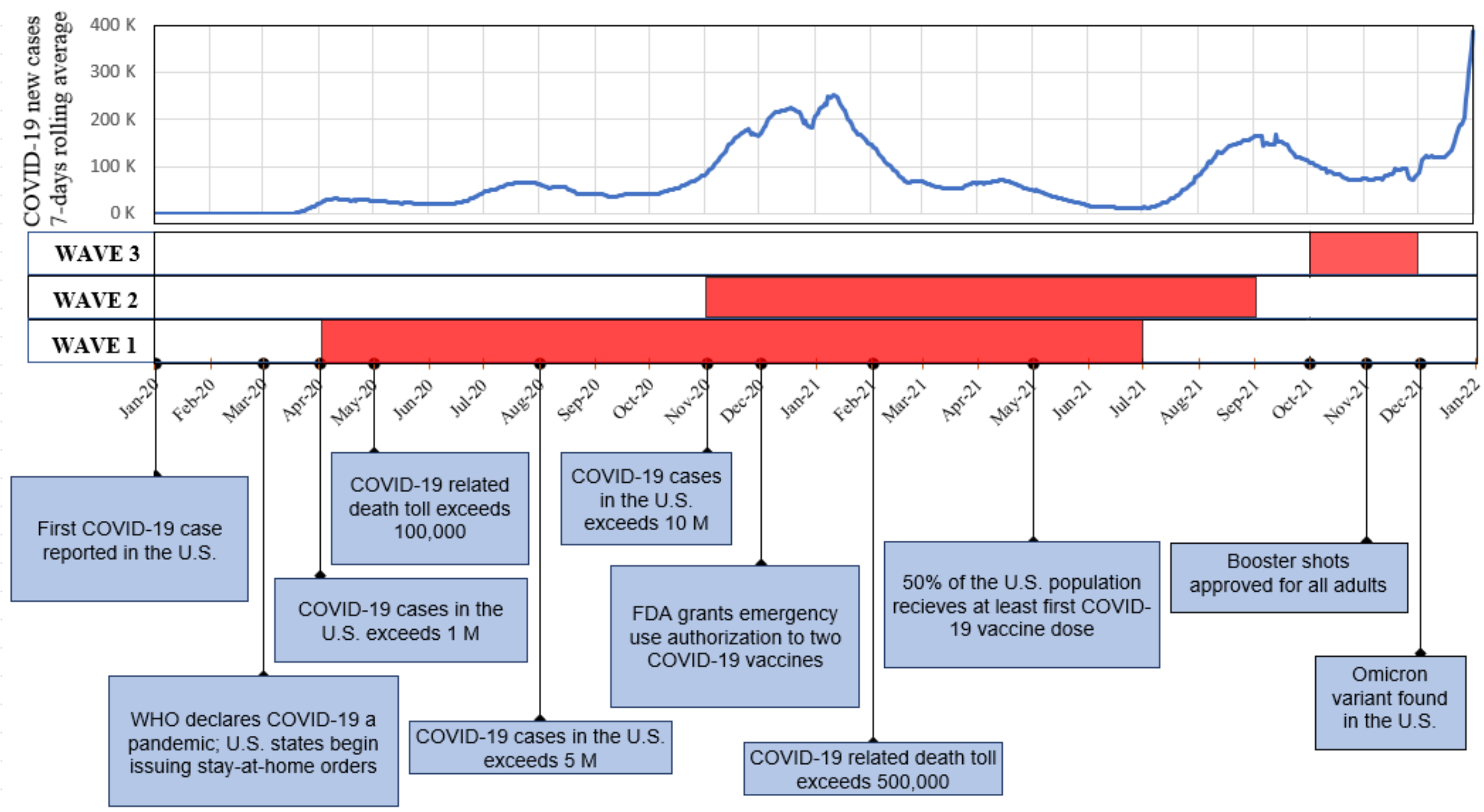

**Figure 1: Timeline of survey waves along with important pandemic related events** (1,2,8–10)

The goal of our survey is to collect data that can provide a better understanding of the public response to COVID-19, and how the responses have evolved over time. This survey dataset can be helpful for government agencies, businesses, and researchers to enable a stronger recovery, to adapt to the changing lifestyles of the American people, and to better prepare to future health crises. This article details the data collection, cleaning, and weighting procedures conducted in the panel survey, with a focus on the weighting procedure from one wave to the next. More general details about the first wave of the survey can be found in Chauhan et al. (11).

A key contribution of the COVID Future study is that we focus primarily on expectations for a post-pandemic world and ask respondents the reasons for any changes they foresee in their behavior to improve the reliability of the responses. In addition, the uniqueness of our panel survey is also observed by containing data that may be analyzed in five periods: pre-pandemic, wave-1, wave-2, wave-3, and post-pandemic period (for which individuals indicated their expectations of behaviors).

The survey dataset is available to download in comma-separated value (CSV) format from the ASU Dataverse (12). Data codebooks and survey instruments can also be downloaded from the ASU Dataverse. It also includes the list of all the survey questions that were asked in each of the survey waves. The data files are versioned sequentially in order of the date of upload. Analyses in this article used version 1.1.0b7 of Wave 1 data, version 2.0.0b6 of Wave 2 data, and version 3.0.0b6 of Wave 3 data.

## SURVEY DESIGN

The survey consisted of multiple-choice questions, Likert scale questions, matrix questions, and open-ended questions. The survey was created on the Qualtrics platform. Figure 2 shows a snapshot of one of the survey questions as seen by the survey participants. The survey was exempt from continuing review by the Institutional Review Board (IRB) offices of the University of Illinois at Chicago (UIC), Arizona State University (ASU), and the University of North Carolina at Chapel Hill (UNC).

The survey questions asked information about a wide variety of topics, including work, studies, shopping and dining, transport, long distance travel, commute to work and/or school, attitudes towards the pandemic, and demographic questions. The first wave asked questions to gather information about the behavior before the beginning of the pandemic (pre-COVID-19), current situation (during COVID-19) and the expected behavior for a time when COVID-19 will no longer be a threat (post COVID-19). Waves 2 and 3 mostly asked questions about during COVID-19 and post COVID-19 period. The questions about pre-COVID-19 were asked only to respondents who did not answer those questions in the first wave of the survey. Similarly, most demographic questions were not asked again, unless respondents skipped that question in the prior wave(s), or respondents reported they have changed. Several key demographics that have been in flux during the pandemic for many, such as employment, were asked again of all respondents.

Although most of the questions were kept similar in all three waves, new questions were added as they became relevant in the context of the pandemic. For instance, the question gauging respondents' risk perception of sending children to school was added in Wave 2 and the questions asking respondents about their trust in the COVID-19 vaccines were added in Wave 3. In a rare situation, we included a new question in between the survey recruitment process. Since the Wave 1 respondents did not have access to the vaccines (outside of clinical trials), the question about their vaccination status was not asked in Wave 1. The vaccines were approved shortly after the Wave 2 began and it took several months for them to be widely available. Therefore, on March 3$^{rd}$, 2021, once vaccines became available to certain groups in the general population, we added three questions to the survey asking about vaccination status and plans. About 55% of (weighted) respondents responded before this date, and did not provide vaccination information, though we assume the vast majority of them were unvaccinated.

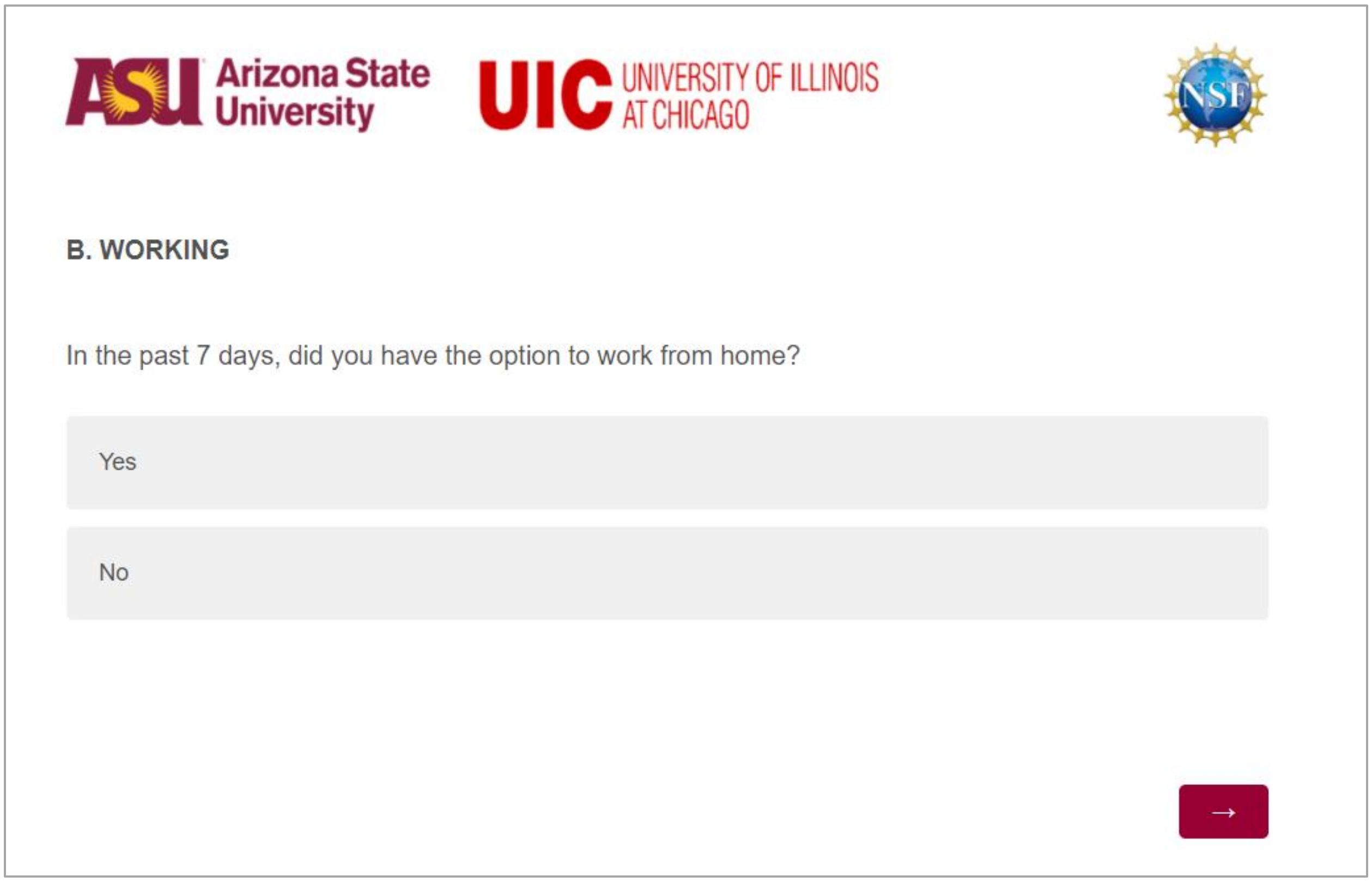

**Figure 2: Snapshot of the online survey**

In addition, in some questions, the respondents were asked "why" questions to provide the reason for their answer. For example, the questions did not only ask how the pandemic had affected the work productivity of the respondents, but it also asked why they thought their work productivity had changed. Similarly, the respondents were asked about: why their learning improved or worsened; why they expect to buy groceries/non-grocery items online more or less; why their household had increased or decreased vehicle ownership; and why they expect to increase/decrease their long-distance travel for personal/leisure purpose, to list a few. These reasons let us dig deeper into the reasons for changes in the future and provide justification for forecasting future changes based on stated preference data.

Waves 2 and 3 took respondents less time to answer, as pre-pandemic behavior and some demographic questions were not asked again. All waves used survey logic so that respondents were only asked the questions that were relevant to them. For instance, an unemployed person would not be asked questions about their current work experience. This resulted the response time varying significantly from respondent to respondent. Table 1 shows the median, 25$^{th}$ percentile, and 75$^{th}$ percentile of response time of each survey wave.

Like the Wave 1 survey, some of the survey logic was based on participants' pre-pandemic behavior. Having respondents re-enter their pre-pandemic behavior would increase respondent burden and likely reduce accuracy compared to Wave 1 since a longer period had passed since the start of the pandemic. Thus, for the Wave 2 and 3 surveys, participants were given a personalized link which used their information from the Wave 1 survey in logic.

**TABLE 1: Median, 25$^{th}$ percentile, and 75$^{th}$ percentile of response time (in minutes) for each survey wave**

| Waves | Median | 25th percentile | 75th percentile |
|---|---|---|---|
| Wave 1 | 21.92 | 16.73 | 30.85 |
| Wave 2 | 16.02 | 12.02 | 23.37 |
| Wave 3 | 19.60 | 14.85 | 28.42 |

**SURVEY RECRUITMENT**

The recruitment for Wave 1 of the survey was conducted through a variety of methods to reach out to a diverse population. These methods include a Qualtrics opinion panel, a purchased list of 450,000 email addresses across the US, a previously purchased list of 39,000 email addresses from the Phoenix metropolitan area, the survey link on the project website, survey invitations sent to family members, friends, and colleagues (also called the convenience sample), and social media advertising. Figure 3 shows the places across the country from where 10,000 or 20,000 emails addresses were purchased, and the cities which were assigned certain sampling quota in Qualtrics. Qualtrics quotas were also assigned for the state of Ohio, Utah, North Carolina, upstate New York, and rural areas. Additionally, a list of 100,000 random email address from across the country (excluding the areas that were already targeted by the remaining 350,000 purchased email addresses) were purchased. Qualtrics quotas were also set to ensure representation from various demographic characteristics, namely age, race and ethnicity, educational attainment, and income. The goal of survey recruitment strategy was to collect responses from across the nation so as to make the data as close to being nationally representative as possible. Details about the recruitment strategy for Wave 1 of the survey can be found in Chauhan et al. (11).

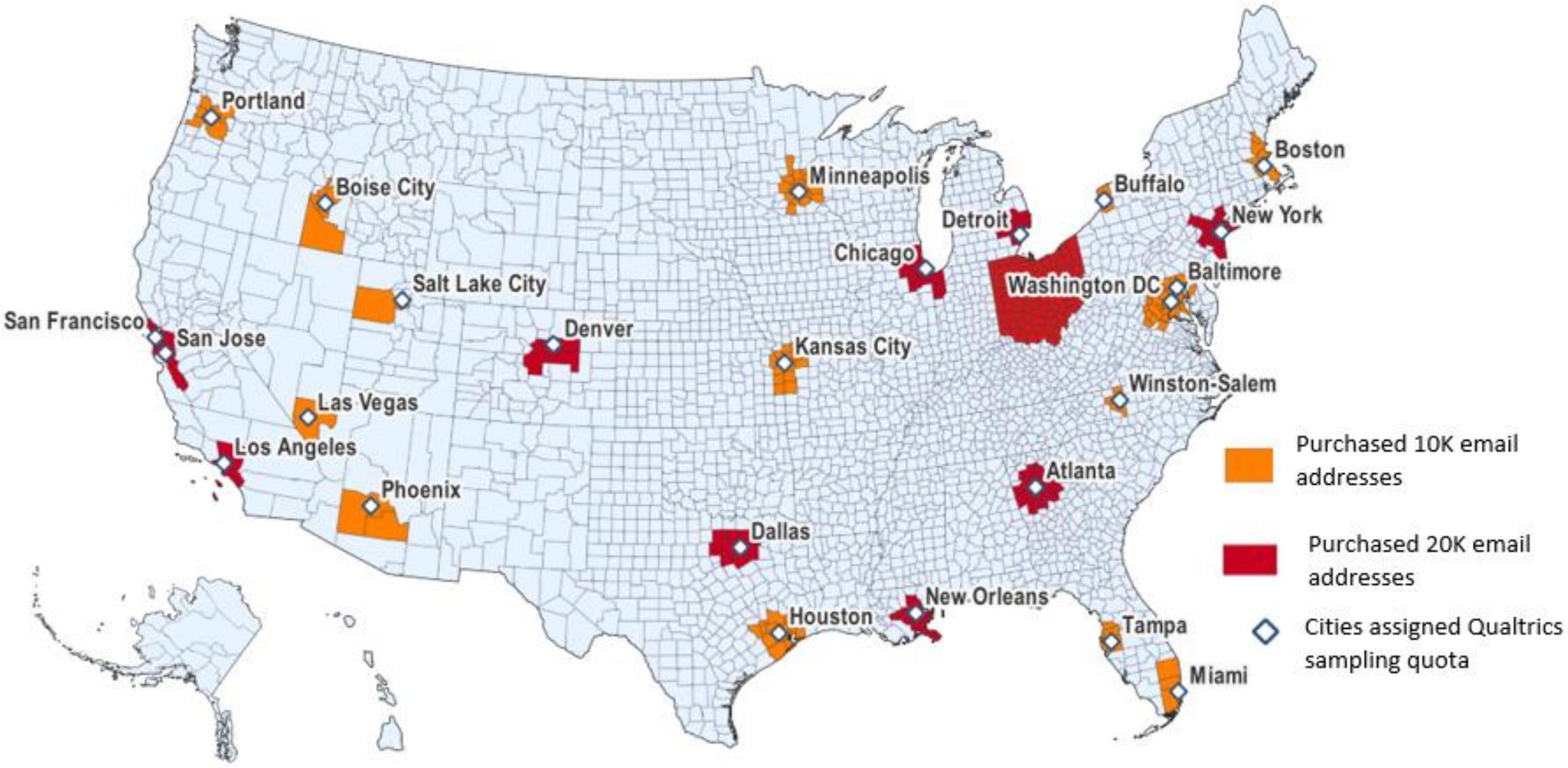

**Figure 3: A county-level map of the United States showing the places targeted by purchased email list and Qualtrics.**

The Wave 1 survey was deployed in two parts. First, an initial survey was deployed to a convenience sample very early in the pandemic, referred to as Wave 1A. A modified and improved version was deployed to purchased email lists and Qualtrics opinion panel in summer and fall 2020, referred to as Wave 1B. In Wave 1, the respondents were asked for their email addresses so that they could be recontacted in subsequent waves.

Only respondents to the Wave 1 survey were invited to participate in Waves 2 and 3. Invitations for Waves 2 and 3 were sent using mass email sending services offered by Amazon Web Services (AWS) and Qualtrics. In addition to providing the survey link, the survey invitation emails included a brief description of the survey, our contact information, and an option to unsubscribe from any further emails. In all three survey waves, all invitees (except for the respondents who responded or unsubscribed our emails) were sent two follow up reminders within a few weeks period to boost the response rate. Since the survey consisted of questions that asked about behavior in the past 7 days, the survey invitations were not sent within 7 days after any major holidays to ensure responses were unaffected by holiday behavior.

Because of our unusual survey collection methodology, the time that elapsed for individual respondents between survey waves differed. Specifically, we left the Wave 1 survey open for nearly a full year, and then sent out Wave 2 invitations 4-6 months after the respondent took the Wave1 survey. Then, we collected Wave 3 all within a one-month period. The time intervals between completing consecutive survey waves are shown is figure 4. The median and average time interval between answering survey waves 1 and 2 were 141 days (or 4.7 months) and 169.4 days (or 5.6 months) respectively. Similarly, the median and average time interval between answering survey waves 2 and 3 were 328 days (or 10.9 months) and 289.2 days (or 9.6 months).

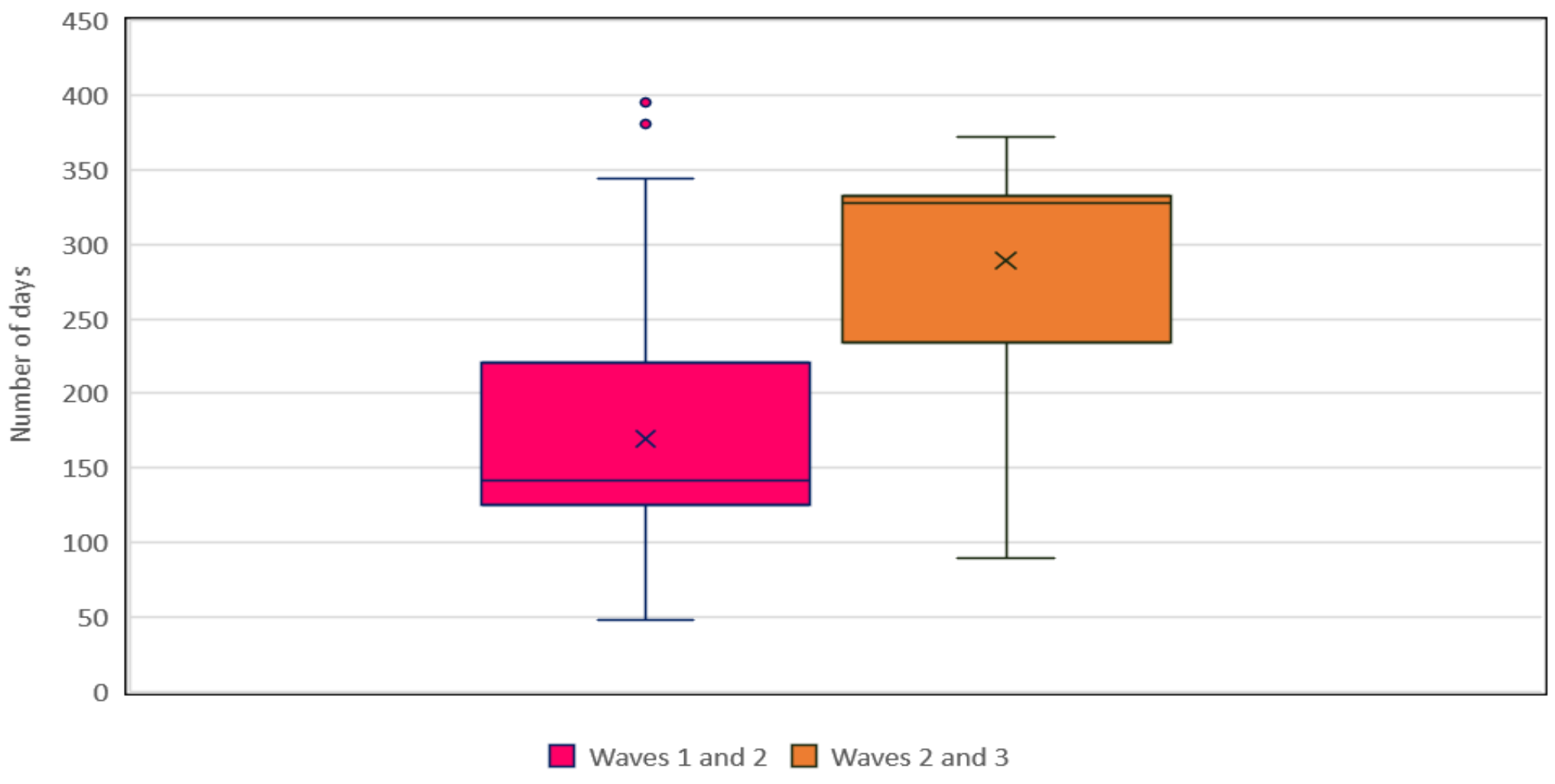

**Figure 4: Time interval in answering consecutive survey waves for respondents who answered all three waves (unweighted, N=1933)**

Incentives were offered to respondents to encourage participation. For Wave 1, a $10 gift card was given to every 20th respondent who was invited through the purchased email addresses. For Waves 2 and 3, all respondents who were not initially recruited through a convenience sample were offered a $5 Amazon gift card upon completion of the follow-up survey.

Prior to taking the survey, the respondents were informed that their participation was voluntary, and that their responses would be shared anonymously on the survey website. The respondents provided online consent to begin the survey. Respondents to Waves 2 and 3 were reminded that they answered the previous survey, and the week during which they answered it. The participants were also shown the partially redacted email address to which the survey was sent to make sure that the survey was answered by the correct person. The participants had the option to start a new survey if they were not the owner of the displayed email address.

Figure 5 shows the sources of data. About 60% of data for each survey wave originated from the Qualtrics online survey panel, followed by those from purchased email list and then by those from the convenience sample. Wave 1 got 7.7% of its respondents from the project website; however, most of these respondents did not stay in the following waves. Figure 6 shows the number of respondents from each U.S. states in each of the survey waves.

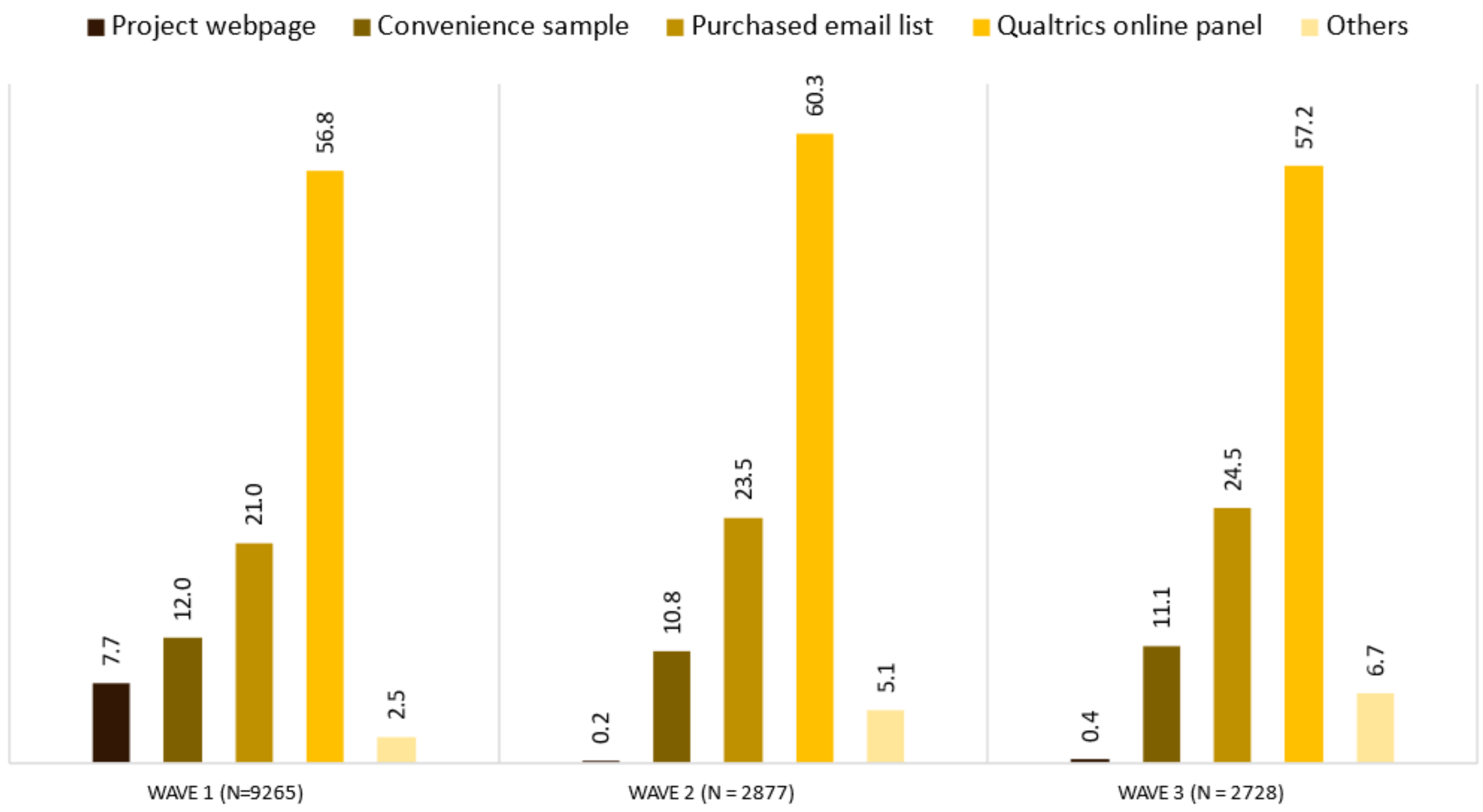

**Figure 5: Source of survey data (in percentages)**

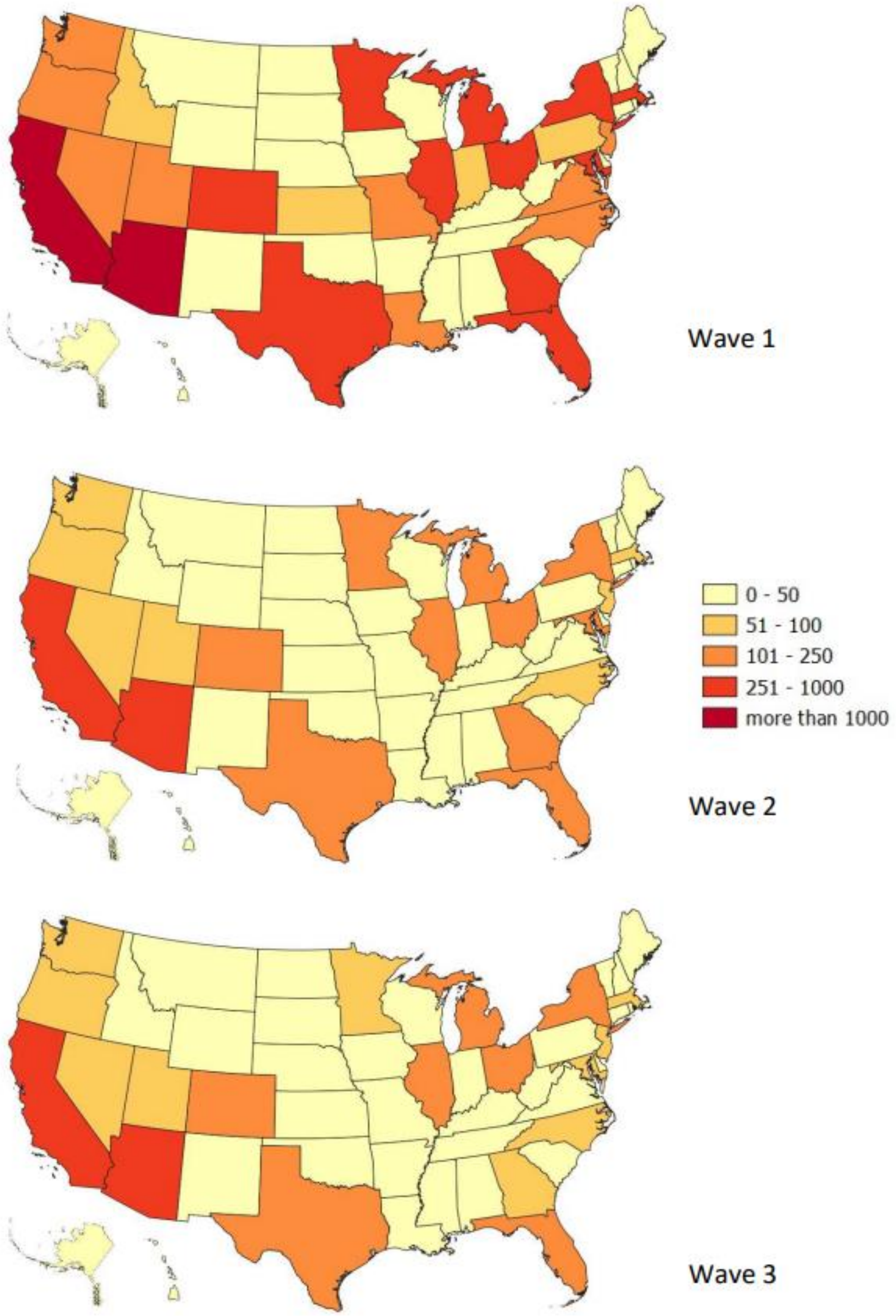

**Figure 6: Survey respondents from each U.S. state in Wave 1 (N=9265), Wave 2 (N=2855), and Wave 3 (N=2708).**

A project website (www.covidfuture.org) was created to disseminate information about the survey. The website also provides the data download link, research findings from the survey dataset, and blogs written by the research team based on the results from survey data analysis. Figure 7 shows a snapshot of the survey website.

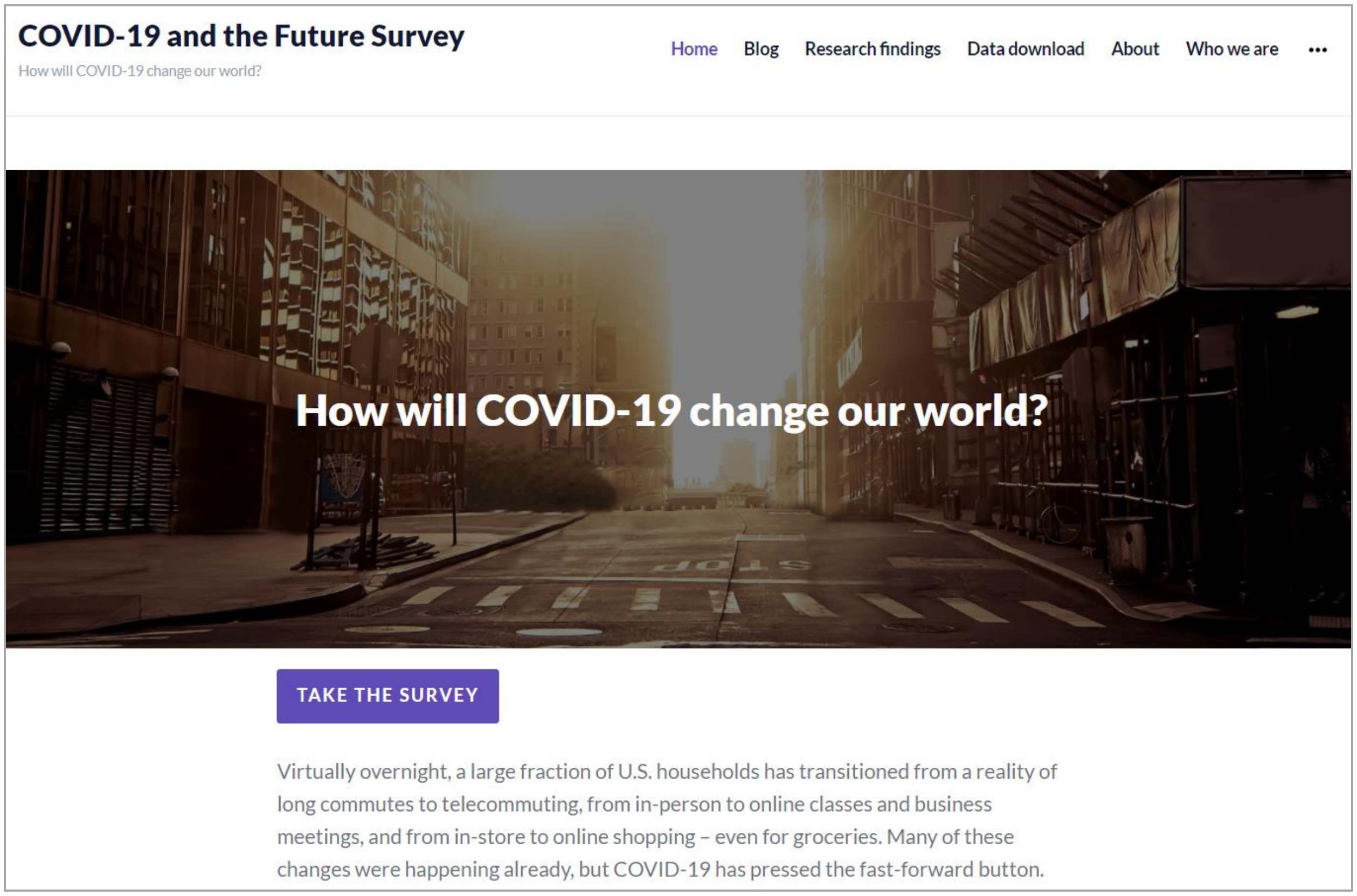

**Figure 7: Snapshot of the survey project website**

## DATA CLEANING

Raw data was cleaned by filtering out any invalid responses to ensure good quality data. For Wave 1, Qualtrics did some attention check and data cleaning. In addition, quality checks were done on the data to remove any illogical responses. More details about Wave 1 data cleaning can be found in Chauhan et al. (11).

Since Wave 2 and Wave 3 were only sent to respondents whose Wave 1 responses were found reliable, there were fewer data quality issues in the Wave 2 and Wave 3 surveys. The primary data cleaning process undertaken for the Wave 2 and Wave 3 surveys was to validate attention check questions and assure proper response time. The Wave 2 and Wave 3 surveys varied in length, and therefore appropriate response time differed. In the Wave 2 survey, all response times were deemed appropriate, with the shortest response time being just under 5 minutes. In the longer Wave 3 survey, any response time which was under five minutes was deemed inappropriate, which resulted in the removal of some survey responses. We included two questions in the shopping and dining, and attitudes sections where respondents were asked to select a particular option to show they were paying attention. This is to improve the response quality (13). If participants missed one of the attention check questions, they were given a second chance to respond to the section where the question appeared, and in the final dataset their responses to these questions were replaced with their second attempt. If they missed both attention check questions once, or either one of them twice, their survey was terminated.

## DATA WEIGHTING

Survey data were weighted to make them nationally representative. A total of 8 sets of weights were developed as shown in table 2.

**Table 2: Weighting variables**

| S.no. | Weight | Description |
|---|---|---|
| 1 | weight_main | Weight for only Wave 1B data. |
| 2 | weight_all_respondents | Weight for all Wave 1 respondents, including Wave 1A. |
| 3 | w2_weight_main | Weights for Wave 2 respondents who originally responded to Wave 1B. |
| 4 | w2_weight_all_respondents | Weights for all Wave 2 respondents, including those who originally responded to Wave 1A |
| 5 | w3_weight_main | Weights for Wave 3 respondents who originally responded to Wave 1B. |
| 6 | w3_weight_all_respondents | Weights for all Wave 3 respondents, including those who originally responded to Wave 1A |
| 7 | w3_w2_weight_main | Weights for all who responded to both Wave 2 and Wave 3, and who originally responded to Wave 1B |
| 8 | w3_w2_weight_all_respondents | Weights for all who responded to both Wave 2 and Wave 3, including those who originally responded to Wave 1A |

Weights weight_main and weight_all_respondents were calculated for Wave 1 data to reflect the overall population of the US, as well as the subpopulation of the 9 regions (Figure 8), roughly corresponding to Census divisions. Data were weighted to be representative in terms of age, education, gender, Hispanic status, household income, presence of children, employment status, white race, household size, tenure status, and number of household vehicles. These weights were computed through iterative proportional fitting (IPF) technique using PopGen2.0 software (14) to match the marginal control totals from the US Census and Public Use Microdata Sample (15–17). Full details of Wave 1 data weighting process can be found in Chauhan et al. (11).

The general practice of weighting the later waves of a panel survey is by adjusting the weights of the first wave (18). The weighting of Waves 2 and 3 of the COVID Future survey is done similarly. An advantage of this weighting process over applying the IPF method again is that the weights it produces are highly correlated across the waves making cross-sectional comparisons more robust.

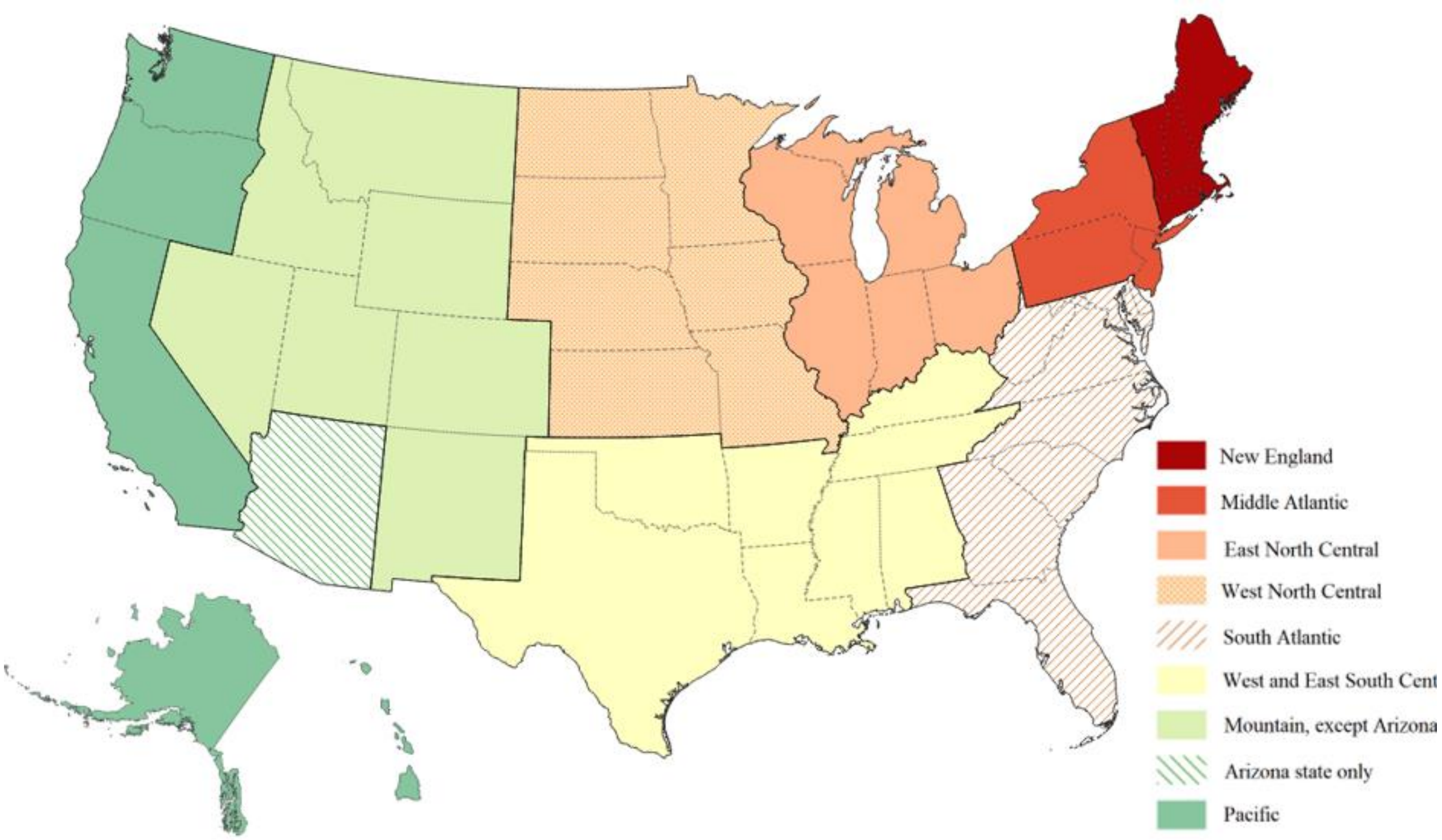

**Figure 8: Wave 1 weighting divisions**

The weighting process of Waves 2 and 3 involves two steps. First, three extensive logistic regression models were estimated to predict nonresponse in Waves 2 and 3. The first model was estimated for Wave 2, the second for Wave 3, and the third for Waves 2 and 3 combined. The third model was needed to determine weights for respondents who answered both Waves 2 and 3. These weights were necessary for the analysis of the intersection of the Waves 2 and 3 subsamples since some of the Wave 3 respondents did not respond to Wave 2. Table 3 shows the results of each of these models. These models predict the probability of answering the survey and not the probability of attrition.

The Wave 2 model has a fair goodness-of-fit, having pseudo-$R^2$ of 0.08. The model results were considerably better for respondents who originally responded to Wave 1B (AUC = 0.72) than those coming from Wave 1A (AUC = 0.55). Whether the lower performance on Wave 1A respondents is due to randomness or unobserved variation is unknown. However, the lower AUC might not be concerning since there is a considerable randomness. The original Wave 1 weights were adjusted for nonresponse by dividing by the predicted probabilities obtained from the logistic regression model. Therefore, the weights of the respondents who were less likely to respond to Wave 2 were inflated.

The pseudo-$R^2$ for Wave 3 model is 0.07, which is similar to that of the model for Wave 2. The pseudo-$R^2$ of the Wave 3 model drastically increases upon adding a variable indicating whether the respondent has answered Wave 2. However, the marginals obtained from that model were further from the control totals. Therefore, it was concluded that the weights from such a model would be of lower quality. Thus, this variable was dropped out of the final Wave 3 model. This brought the performance of the Wave 3 combined model in line with the other two regression models.

Each set of weights (Wave 2, Wave 3, and Waves 2 and 3 combined) were also calculated separately for respondents who responded to Wave 1B (i.e., Wave 1 excluding Wave 1A). The weights for respondents who responded to Wave 1B are referred to as 'main' while the respondents who responded to Wave 1 (including Wave 1A) are referred to as 'all respondents.' We did not fit separate logistic regression models for the Wave 1B subsample, but rather used the predicted probabilities from the full-sample models to adjust the original weights developed for the full Wave 1B subsample. The separate weighting for Wave 1B respondents was deemed necessary because there were several questions that were changed from Wave 1A to Wave 1B. Additionally, Wave 1A is likely biased since all its respondents were recruited through a convenience sample. Therefore, we recommend most data users use the 'main' subsample.

The work-from-home variables are arguably the most critical variables of the survey. Therefore, the second step of the weighting process was to ensure that distribution of people working from home before the pandemic, during Wave 1, and post pandemic as they predicted at the time of Wave 1 matched exactly between all waves. This would confirm that any differences between the waves in current or expected work from home outcomes are not due to sample selection. Henceforth, marginal distributions of the work-from-home variables were created based on the Wave 1 dataset and separately for Wave 1B dataset. Next, the weights from the logistic regression nonresponse adjustment were used to create a seed matrix for an IPF with work-from-home based marginals. This IPF slightly adjusts the logistic regression-based weights to perfectly match Wave 1 or Wave 1B on these critical variables.

Further, the correlations were determined between all the different sets of weights. For Wave 2, it was found that the base logit-adjusted weights have a correlation of 0.99 with the final weights obtained after including the additional IPF step. This correlation ranged from 0.94 to 0.98 in Wave 3. These extremely high correlations indicate that while the additional step of including an IPF improved the accuracy of the work-from-home outcomes, the weights were only slightly adjusted. Thus, they retain the advantages of the original regression-based weights.

Table 4 shows the comparison between the distribution of key demographic characteristics in the weighted survey data and the ACS 2018 1-year estimates. The weighted data distributions closely match with the distributions in the ACS 2018 data with some small unavoidable variations. One notable difference was that the weighted data distribution slightly overestimates the population between age 45 - 64 years, and slightly underestimates the population of age 65 years and older.

**Table 3: Results from the nonrespondent model for Wave 2.**

| | Wave 2 | | Wave 3 | | Waves 2 and 3 | |
|---|---|---|---|---|---|---|
| | OR | p-value | OR | p-value | OR | p-value |
| Constant | 0.19*** | 0.000 | 0.238*** | 0.000 | 0.116*** | 0.000 |
| Non-Hispanic and Black/African American | 0.645*** | 0.000 | 0.856 | 0.105 | 0.647*** | 0.000 |
| Non-Hispanic and Asian | 1.15 | 0.191 | 1.078 | 0.495 | 1.083 | 0.508 |
| Non-Hispanic and American Indian and Alaska Native | 0.783 | 0.292 | 0.896 | 0.636 | 0.858 | 0.563 |
| Non-Hispanic and Native Hawaiian or | 0.469 | 0.172 | 0.488 | 0.195 | 0.406 | 0.227 |
| Other Pacific Islander | | | | | | |
| Non-Hispanic and some other race than those mentioned above or White | 0.901 | 0.598 | 0.874 | 0.502 | 0.998 | 0.994 |
| Child 4 or younger in household | 0.972 | 0.769 | 1.008 | 0.932 | 1.027 | 0.812 |
| Child 5–12 in household | 0.796** | 0.004 | 0.803** | 0.006 | 0.714*** | 0.000 |
| Child 13–17 in household | 0.894 | 0.156 | 0.899 | 0.187 | 0.88 | 0.165 |
| Homeowner | 1 | 0.999 | 1.02 | 0.770 | 1.036 | 0.635 |
| Had COVID | 0.628* | 0.041 | 0.47** | 0.002 | 0.594 | 0.065 |
| Lives in single-family home | 1.016 | 0.811 | 1.038 | 0.586 | 1.022 | 0.777 |
| Would like to continue some pandemic lifestyle changes | 1.152* | 0.036 | 1.069 | 0.329 | 1.107 | 0.183 |
| Maybe would like to continue some COVID lifestyle changes | 1.094 | 0.143 | 1.112 | 0.087 | 1.129 | 0.081 |
| Had option to work from home pre-pandemic | 1.021 | 0.897 | 0.816 | 0.236 | 0.883 | 0.518 |
| Had option to work from home during Wave 1 | 1.054 | 0.586 | 1.134 | 0.200 | 1.107 | 0.356 |
| Expected to have option to work from home after the pandemic | 1.035 | 0.914 | 1.157 | 0.648 | 1.104 | 0.786 |
| Likes working from home (binary) | 0.868** | 0.010 | 0.872* | 0.013 | 0.858* | 0.013 |
| Wave 1A respondent recruited via email | 0.931 | 0.634 | 1.001 | 0.995 | 0.932 | 0.679 |
| Home ZIP code denser than 2000 housing units per square kilometer | 1.031 | 0.715 | 1.046 | 0.600 | 1.039 | 0.683 |
| Household income: $100,000 or more | 0.935 | 0.277 | 0.891 | 0.061 | 0.864* | 0.034 |
| Household income: Less than $35,000 | 0.986 | 0.842 | 0.779*** | 0.001 | 0.813* | 0.011 |
| Region: East North Central | 1.825*** | 0.000 | 1.409*** | 0.000 | 1.928*** | 0.000 |
| Region: Mid-Atlantic | 1.507*** | 0.000 | 1.179 | 0.148 | 1.608*** | 0.000 |
| Region: Mountain, except Arizona | 1.379** | 0.002 | 1.045 | 0.667 | 1.342* | 0.014 |
| Region: New England | 1.589*** | 0.001 | 1.257 | 0.104 | 1.507* | 0.011 |
| Region: Pacific | 1.404*** | 0.000 | 1.197 | 0.056 | 1.578*** | 0.000 |
| Region: South Atlantic | 1.526*** | 0.000 | 1.139 | 0.168 | 1.595*** | 0.000 |
| Region: West North Central | 1.613*** | 0.000 | 1.225 | 0.101 | 1.681*** | 0.000 |
| Region: West/East South Central | 1.436*** | 0.001 | 1.05 | 0.658 | 1.424** | 0.005 |
| Employment status now: No, I'm furloughed with pay | 0.884 | 0.627 | 0.882 | 0.626 | 0.738 | 0.312 |
| Employment status now: No, I'm furloughed without pay | 1.058 | 0.711 | 1.026 | 0.865 | 0.976 | 0.890 |
| Employment status now: Yes, working full-time | 1.066 | 0.625 | 0.914 | 0.501 | 0.895 | 0.460 |
| Employment status now: Yes, working part-time (or reduced hours in Wave 1B) | 1.039 | 0.766 | 0.95 | 0.702 | 0.964 | 0.804 |
| Employment status before the pandemic: Employed full time | 0.752* | 0.019 | 0.814 | 0.097 | 0.881 | 0.366 |
| Employment status before the pandemic: Employed part time | 0.895 | 0.381 | 0.963 | 0.772 | 0.978 | 0.878 |
| Were you a student in Spring 2020?: Yes, Full-time | 0.82 | 0.073 | 0.868 | 0.211 | 0.759* | 0.039 |
| Were you a student in Spring 2020?: Yes, Part-time | 0.696** | 0.008 | 0.556*** | 0.000 | 0.559*** | 0.001 |
| What is your educational background?: Some grade/high school | 0.528* | 0.011 | 0.683 | 0.116 | 0.559 | 0.058 |
| What is your educational background?: Some college or technical school | 1.326*** | 0.001 | 1.166 | 0.070 | 1.211* | 0.044 |
| What is your educational background?: Bachelor's degree(s) or some graduate school | 1.65*** | 0.000 | 1.558*** | 0.000 | 1.616*** | 0.000 |
| What is your educational background?: Completed graduate degree(s) | 1.522*** | 0.000 | 1.47*** | 0.000 | 1.529*** | 0.000 |
| gender: Female | 1.323*** | 0.000 | 1.186*** | 0.001 | 1.223*** | 0.000 |
| Do you consider yourself Hispanic or Latino/a?: Yes | 1.006 | 0.978 | 1.159 | 0.498 | 0.901 | 0.669 |
| Household vehicles: 0 | 1.242* | 0.035 | 1.186 | 0.105 | 1.328* | 0.014 |
| Household vehicles: 2 | 1.075 | 0.216 | 0.963 | 0.525 | 1.016 | 0.806 |
| Household vehicles: 3 | 0.865 | 0.095 | 0.809* | 0.016 | 0.847 | 0.096 |
| Household vehicles: 4 or more | 0.883 | 0.309 | 0.971 | 0.809 | 0.935 | 0.635 |
| After COVID-19 is no longer a threat, how often might you work from home?: A few time/year | 0.707 | 0.328 | 0.692 | 0.297 | 0.649 | 0.287 |
| After COVID-19 is no longer a threat, how often might you work from home?: A few times/month | 1.084 | 0.804 | 0.889 | 0.718 | 1.028 | 0.941 |
| After COVID-19 is no longer a threat, how often might you work from home?: A few time/week | 1.155 | 0.659 | 0.907 | 0.763 | 1.035 | 0.927 |
| After COVID-19 is no longer a threat, how often might you work from home?: Every day | 1.127 | 0.724 | 0.874 | 0.691 | 1.009 | 0.981 |
| Before the COVID-19 pandemic, how often did you usually work from home?: A few times/year | 0.905 | 0.600 | 1.265 | 0.230 | 1.093 | 0.687 |
| Before the COVID-19 pandemic, how often did you usually work from home?: A few times/month | 0.821 | 0.309 | 1.022 | 0.914 | 0.851 | 0.480 |
| Before the COVID-19 pandemic, how often did you usually work from home?: | 1.089 | 0.670 | 1.265 | 0.260 | 0.984 | 0.945 |
| Once/week | | | | | | |
| Before the COVID-19 pandemic, how often did you usually work from home?: A few times/week | 0.809 | 0.285 | 1.099 | 0.647 | 0.925 | 0.736 |
| Before the COVID-19 pandemic, how often did you usually work from home?: Every day | 1.099 | 0.650 | 1.449 | 0.087 | 1.198 | 0.456 |
| Recruited from purchased nationwide email list | 0.879 | 0.069 | 0.897 | 0.128 | 0.676*** | 0.000 |
| Respondent not recruited through InfoGroup/Qualtrics (convenience sample or referral from other respondent) | 0.27*** | 0.000 | 0.44*** | 0.000 | 0.296*** | 0.000 |
| Recruited from convenience sample | 0.66*** | 0.000 | 0.728** | 0.002 | 0.679*** | 0.001 |
| Response time 30-60 minutes | 0.816** | 0.001 | 0.756*** | 0.000 | 0.8** | 0.002 |
| Response time >60 minutes | 0.977 | 0.800 | 0.894 | 0.229 | 0.974 | 0.800 |
| Age 30-44 | 1.737*** | 0.000 | 1.892*** | 0.000 | 2.09*** | 0.000 |
| Age 44-59 | 2.305*** | 0.000 | 2.423*** | 0.000 | 2.422*** | 0.000 |
| Age 60+ | 2.489*** | 0.000 | 2.844*** | 0.000 | 2.747*** | 0.000 |
| Whether the user responded to the survey on a mobile device (based on operating system): Yes | 0.665*** | 0.000 | 0.721*** | 0.000 | 0.722*** | 0.000 |
| How would a $400 emergency expense that you had to pay impact your ability to pay your other bills this month?: Seen but unaswered | 1.083 | 0.852 | 1.629 | 0.216 | 0.714 | 0.544 |
| How would a $400 emergency expense that you had to pay impact your ability to pay your other bills this month?: I don't know | 0.653*** | 0.000 | 0.709*** | 0.000 | 0.59*** | 0.000 |
| How would a $400 emergency expense that you had to pay impact your ability to pay your other bills this month?: I prefer not to answer | 0.665* | 0.027 | 0.536** | 0.002 | 0.591* | 0.020 |
| How would a $400 emergency expense that you had to pay impact your ability to pay your other bills this month?: I would not be able to pay all my bills | 0.711*** | 0.000 | 0.683*** | 0.000 | 0.65*** | 0.000 |
| Pseudo-R2 | 0.077472 | | 0.071408 | | 0.077208 | |
| *: *p* < 0.05, **: *p* < 0.01, ***: *p* < 0.001 | | | | | | |

**Table 4: Comparison of unweighted and control distributions of demographics**

| | | weight | weig | w2_weig | w2_w | w3_weig | w3_w | w3_w2_w | w3_w2 | ACS |
|---|---|---|---|---|---|---|---|---|---|---|

| | | _all_respondents | ht_main | ht_all_respondents | eight_main | ht_all_respondents | eight_main | eight_all_respondents | _weight_main | 2018 1-year estimates |
|---|---|---|---|---|---|---|---|---|---|---|
| Number of respondents (N) | | 9265 | 8152 | 2877 | 2566 | 2728 | 2424 | 1933 | 1716 | - |
| Person Characteristics | | | | | | | | | | |
| Age | 18–29 years | 20.9% | 20.9% | 18.3% | 17.3% | 18.6% | 17.3% | 17.1% | 15.8% | 21.2% |
| | 30–44 years | 25.2% | 25.3% | 27.0% | 26.7% | 26.7% | 26.4% | 28.6% | 28.1% | 25.1% |
| | 45–64 years | 34.1% | 33.8% | 34.0% | 34.8% | 33.3% | 34.3% | 34.6% | 35.7% | 24.9% |
| | 65 years and above | 19.7% | 20.0% | 20.8% | 21.3% | 21.3% | 22.0% | 19.7% | 20.3% | 28.8% |
| Gender | Male | 48.7% | 48.7% | 48.4% | 47.8% | 49.1% | 49.1% | 46.8% | 46.1% | 48.7% |
| | Female | 51.3% | 51.3% | 51.6% | 52.2% | 50.9% | 50.9% | 53.2% | 53.9% | 51.3% |
| Education | High School Degree or Less | 39.0% | 39.0% | 36.5% | 35.8% | 36.5% | 36.1% | 35.9% | 35.0% | 39.3% |
| | Some College or Associate’s Degree | 30.4% | 30.4% | 31.5% | 32.0% | 30.7% | 30.5% | 30.9% | 30.9% | 30.6% |
| | Bachelor’s Degree or Higher | 30.6% | 30.6% | 32.1% | 32.3% | 32.8% | 33.5% | 33.3% | 34.1% | 30.1% |
| Employment | Employed | 62.0% | 62.0% | 62.0% | 62.0% | 56.3% | 62.0% | 58.9% | 62.0% | 61.5% |
| | Not employed | 38.0% | 38.0% | 38.0% | 38.0% | 43.7% | 38.0% | 41.1% | 38.0% | 38.5% |
| Ethnicity | Hispanic | 16.4% | 16.4% | 15.0% | 14.6% | 14.8% | 14.6% | 15.7% | 15.4% | 16.2% |
| | Not Hispanic | 83.6% | 83.6% | 85.0% | 85.4% | 85.2% | 85.4% | 84.3% | 84.6% | 83.8% |
| Race | White | 73.6% | 73.6% | 77.1% | 77.0% | 75.6% | 75.5% | 78.4% | 78.3% | 73.8% |
| | Non-White | 26.4% | 26.4% | 22.9% | 23.0% | 24.4% | 24.5% | 21.6% | 21.7% | 26.2% |
| Person Level Controls of Household Characteristics | | | | | | | | | | |
| Size | 1 | 16.7% | 16.7% | 17.7% | 17.6% | 15.4% | 15.3% | 16.0% | 16.0% | 16.5% |
| | 2 | 32.9% | 32.9 % | 33.9% | 33.9 % | 33.6% | 34.0 % | 33.0% | 33.4% | 32.8% |
| | 3 | 18.7% | 18.7 % | 18.2% | 18.1 % | 22.5% | 22.4 % | 21.0% | 20.6% | 19.0% |
| | 4 or larger | 31.7% | 31.7 % | 30.1% | 30.4 % | 28.5% | 28.2 % | 30.0% | 30.0% | 31.7% |
| Presence of children | Present | 32.8% | 32.8 % | 32.3% | 32.9 % | 33.9% | 34.3 % | 33.8% | 34.8% | 35.9% |
| | Not present | 67.1% | 67.1 % | 67.7% | 67.1 % | 66.1% | 65.7 % | 66.2% | 65.2% | 64.1% |
| Tenure | Homeowner | 65.6% | 65.7 % | 66.1% | 67.1 % | 66.3% | 67.4 % | 65.2% | 66.4% | 65.4% |
| | Not homeowner | 34.3% | 34.3 % | 33.8% | 32.9 % | 33.7% | 32.5 % | 34.8% | 33.6% | 34.6% |
| Vehicles available | 0 | 9.3% | 9.3% | 9.4% | 9.2% | 9.2% | 9.1% | 8.9% | 8.9% | 9.3% |
| | 1 | 22.6% | 22.6 % | 24.6% | 24.0 % | 22.5% | 22.0 % | 23.0% | 22.2% | 22.8% |
| | 2 | 37.4% | 37.4 % | 37.4% | 36.9 % | 37.8% | 37.7 % | 36.7% | 35.8% | 37.5% |
| | 3 or more | 30.7% | 30.7 % | 28.7% | 29.8 % | 30.4% | 31.1 % | 31.5% | 33.1% | 30.4% |
| 2019 income before taxes | Less than $35,000/year | 18.9% | 18.9 % | 18.6% | 18.4 % | 17.8% | 17.6 % | 17.5% | 17.1% | 20.7% |
| | $35,000 to $99,999/year | 41.1% | 41.1 % | 39.7% | 39.4 % | 40.1% | 40.3 % | 39.2% | 39.0% | 41.8% |
| | More than $100,000/year | 40.0% | 40.0 % | 41.7% | 42.2 % | 42.1% | 42.1 % | 43.2% | 43.9% | 37.5% |

**RESULTS AND DISCUSSION**

The COVID Future survey captures the attitudes and behaviors during the different phases of the pandemic. The Wave 1 survey was deployed at the height of the pandemic, with the bulk of data collection occurring between June and October of 2020. Wave 2, on the other hand, was deployed during the transition phase of the pandemic. When the Wave 2 survey recruitment began in November 2020, the pandemic was still severe; but the COVID-19 cases kept dropping during most of the Wave 2 data collection (Figure 1). By the completion of Wave 2 in August 2021, vaccination was widely available in the US, and states were gradually lifting COVID-19 restrictions. Thus, during the Wave 2 data collection and especially towards the end of data collection, life in the US was significantly closer to "normal" than it was during Wave 1 data collection. When the Wave 3 was conducted during October – November 2021, it had been around 1.5 years since the pandemic began and hence the pandemic had become more-or-less a part of daily life. By this time, most of the COVID-19 restrictions, including those on travel, mass gathering, were lifted. Additionally, it was during the Wave 3 data collection when booster shots were approved for all adults in the US (figure 1). Therefore, the time period of Wave 3 was of further stability and normalcy than Wave 2.

Since the three waves of the COVID Future survey were conducted in different phases of the pandemic, the survey data reflects the evolution of lifestyle and thinking through the pandemic. In both Waves 1 and 2, 57% of the respondents were concerned about getting severe reaction from coronavirus if they catch it, however, only 42% reported the same in Wave 3. The concern about friends and family getting severe reaction from coronavirus was reported by 76% in Wave 1, 74% in Wave 2, but only 61% in Wave 3. The views on staying-at-home and shutting down businesses also changed over time. In Waves 1 and 2, about 75% respondents favored staying at home as much as possible until the coronavirus has subsided. This percentage dropped to 46% in Wave 3. Similarly, 57% of respondents in Wave 1 supported shutting down business to prevent the spread of coronavirus. This percentage declined to 49% in Wave 2, and only 40% in Wave 3. The survey data also shows how the perspective on social behavior evolved through the pandemic. While 50% of respondents viewed video calling as a good alternative to visiting friends and family in Wave 1, this percentage dropped to 47% in Wave 2 and got to merely 38% in Wave 3. However, around 68% of respondents saw video calling as a good alternative to in-person business meetings – in all three waves. The survey data also captured the change in opinions on online learning. 48% of Wave 1 respondents saw online learning as a good alternative to high school- and college-level classroom instruction. This initial support got reduced to 46% in Wave 2 and 41% in Wave 3.

The survey data also captures the changes in respondents' risk perception through the pandemic. In Wave 1, 26% of respondents viewed high or extremely high risk in shopping at grocery store while the same was reported by 23% in Wave 2 and only 14% in Wave 3. High risk or extremely high risk in riding in public transit was reported by 71% of Wave1 respondents, 65% of Wave 2 respondents, and 52% of Wave 3 respondents. High or extremely high risk in taking a taxi or ride hailing service was reported by 44% of Wave 1 respondents, 36% of Wave 2 respondents, and 23% of Wave 3 respondents. Similarly, high risk or extremely high risk in traveling in airplane was viewed by 69% of Wave 1 respondents, 59% of Wave 2 respondents, and 45% of Wave 3 respondents. These changes in attitudes and risk perceptions likely translate to increased comfort engaging in social activity.

The initial shock in response to the pandemic and the gradual return to 'pre-covid normal' over the course of the pandemic are both reflected in the data. Before the beginning of the pandemic, 34% of respondents were using public transit, however, only 7% reported the use of public transit within last 7 days of the survey in Wave 1. This number increased to 8% in Wave 2 and 12% in Wave 3. Similarly, 37% respondents took taxi or ride hailing services before the pandemic. Though only 5% in Wave 1 reported doing the same in past 7 days of survey. This number increased slightly to 6% in Wave 2 and 9% in Wave 3. We did not ask about during-pandemic air travel in Wave 1 since air travel was so rare at that time. However, air travel had rebounded somewhat by the time Wave 2 was conducted, with 13% of respondents traveling by air in the three months preceding the survey. This number jumped to 31% in Wave 3. As predicted based on Wave 1 data, personal travel has rebounded more quickly than leisure travel (11,19,20). 11% of respondents reported leisure travel while only 3% reported business travel in Wave 2; and 30% of respondents reported leisure travel versus merely 5% reported business travel in Wave 3.

Over the course of the pandemic, the respondents were more likely to travel to potentially crowded places. In Wave 1, 79% of respondents reported shopping for groceries in a store at least once in the last 7 days. This percentage grew to 83% in Wave 2 and 88% in Wave 3. Similarly, only 20% of respondents dined in at a restaurant at least once in the last seven days in Wave 1 versus 29% in Wave 2 and as much as 53% in Wave 3.

Working from home, however, was prevalent through the pandemic, with 56% of the employed respondents in Wave 3 reported having the option to work from home which remained essentially unchanged from 57% in Wave 1 and 51% in Wave 2.

These differences in context between the three survey waves affect research in several ways. First, any analysis of ‘during-pandemic’ behaviors may show significantly different results between Waves 1, 2, and 3 as, generally speaking, the later waves took place once restrictions began to be lifted and many respondents were more comfortable in social situations. Second, survey respondents’ expectations for the future are likely to be somewhat more accurate in the later waves since the current lives are beginning to adjust to a “new normal.”

The survey data can assist in policy making. Specific policy-relevant results of the survey can be found in other publications (19–24).

## CONCLUSION

Data-hungry techniques like Machine learning and Deep Learning have become more popular to conduct research (25–28). This trend has made data crucial and a central component of many analyses. Surveys are an effective way to collect data. However, conducting a survey can be complex and challenging. To guide through this process, this article presents a comprehensive summary of all the steps involved in conducting the online, nationwide, longitudinal COVIDFuture survey.

The COVIDFuture survey was conducted to observe the shifts in attitudes and behaviors over the course of the COVID-19 pandemic. It asked questions about work, studies, shopping and dining, transport, long distance travel, commute to work and/or school, attitudes towards the pandemic, and demographic information. This uniquely designed survey was conducted in three waves and presents information about five timeframes: before the pandemic, during survey waves 1, 2, and 3, and the expectations for the post-pandemic period.

This article explained the survey’s design, describing the questions asked in each of the survey waves. It elaborated on the several strategies employed for survey recruitment. It discussed the steps that were deemed necessary for data cleaning and curating. Specifically, it detailed the data weighting methodology for each of the survey wave to make the data nationally representative. Finally, some of key findings from the survey data and their interpretation in the context of the pandemic situation were discussed. Overall, the article communicates the unique challenges that were faced in conducting this survey and the strategies that were used to overcome them.

This article can serve as a step-by-step comprehensive guide on conducting a survey and curating survey data. The methodology explained here can provide important lessons to the future surveys.

## ACKNOWLEDGMENTS

This research was supported in part by the National Science Foundation (NSF) RAPID program under grants no. 2030156 and 2029962 and by the Center for Teaching Old Models New Tricks (TOMNET), a University Transportation Center sponsored by the U.S. Department of Transportation through grant no. 69A3551747116, as well as by the Knowledge Exchange for Resilience at Arizona State University. This COVID-19 Working Group effort was also supported by the NSF-funded Social Science Extreme Events Research (SSEER) network and the CONVERGE facility at the Natural Hazards Center at the University of Colorado Boulder (NSF Award #1841338) and the NSF CAREER award under grant no. 155173.

## AUTHOR CONTRIBUTIONS

The authors confirm contribution to the paper as follows: study conception: DS, RMP, AKM, and SD; study design: RSC, MWBC, TM, ER, DS, RMP, AKM, and SD; data collection: All authors; analysis and interpretation of results: RSC, MWBC, TM, NC, ER, and AD; draft manuscript preparation: RSC, MWBC, TM, and NC. All authors reviewed the results and approved the final version of the manuscript.

## REFERENCES

1. Staff A. A Timeline of COVID-19 Developments in 2020. Am J Manag Care. 2021;1.

2. United States - COVID-19 Overview - Johns Hopkins [Internet]. Johns Hopkins Coronavirus Resource Center. [cited 2020 Nov 11]. Available from: https://coronavirus.jhu.edu/region/united-states

3. USC Dornsife - Understanding Coronavirus in America | Understanding America Study [Internet]. [cited 2022 Jul 19]. Available from: https://covid19pulse.usc.edu/

4. About The Study [Internet]. Post Covid-19 Mobility. 2021 [cited 2022 Jul 19]. Available from: https://postcovid19mobility.sf.ucdavis.edu/about-study

5. Drummond J, Hasnine MS. Online and In-store Shopping Behavior During COVID-19 Pandemic: Lesson Learned from a Panel Survey in New York City. In 2022 [cited 2022 Jul 19]. Available from: https://trid.trb.org/view/1900894

6. Young M, Soza-Parra J, Circella G. The Growth in Online Shopping Frequency During COVID-19: Who is Responsible and is this Increase Temporary or Long-lasting? In 2022 [cited 2022 Jul 19]. Available from: https://trid.trb.org/view/1900891

7. Impact of the COVID-19 pandemic on travel behavior in Istanbul: A panel data analysis. Sustain Cities Soc. 2021 Feb 1;65:102619.

8. Ritchie H, Mathieu E, Rodés-Guirao L, Appel C, Giattino C, Ortiz-Ospina E, et al. Coronavirus Pandemic (COVID-19). Our World Data [Internet]. 2020 Mar 5 [cited 2022 Jul 19]; Available from: https://ourworldindata.org/coronavirus

9. Interim public health recommendations for fully vaccinated people [Internet]. [cited 2022 Jul 19]. Available from: https://stacks.cdc.gov/view/cdc/105629

10. A Timeline of COVID-19 Vaccine Developments for the Second Half of 2021 [Internet]. AJMC. [cited 2022 Jul 19]. Available from: https://www.ajmc.com/view/a-timeline-of-covid-19-vaccine-developments-for-the-second-half-of-2021

11. Chauhan R, Conway M, Capasso da Silva D, Salon D, Shamshiripour A, Rahimi E, et al. A Database of Travel-Related Behaviors and Attitudes Before, During, and After COVID-19 in the United States. Sci Data. 2021;

12. Salon D, Conway MW, Capasso da Silva D, Chauhan R, Shamshiripour A, Rahimi E, et al. COVID Future Wave 1 Survey Data v1.0.0 [Internet]. DRAFT VERSION. ASU Library Research Data Repository; 2020. Available from: https://doi.org/10.48349/ASU/QO7BTC

13. Shamon H, Berning C. Attention Check Items and Instructions in Online Surveys: Boon or Bane for Data Quality? Surv Res Methods Forthcom. 2019;

14. PopGen [Internet]. MARG - Mobility Analytics Research Group. [cited 2020 Nov 11]. Available from: https://www.mobilityanalytics.org/popgen.html

15. Ye X, Konduri K, Pendyala RM, Sana B, Waddell P. A methodology to match distributions of both household and person attributes in the generation of synthetic populations. In: 88th Annual Meeting of the Transportation Research Board, Washington, DC. 2009.

16. Konduri KC, You D, Garikapati VM, Pendyala RM. Enhanced synthetic population generator that accommodates control variables at multiple geographic resolutions. Transp Res Rec. 2016;2563(1):40–50.

17. Bar-Gera H, Konduri KC, Sana B, Ye X, Pendyala RM. Estimating Survey Weights with Multiple Constraints Using Entropy Optimization Methods. In 2009 [cited 2022 Jul 19]. Available from: https://trid.trb.org/view/881144

18. Vandecasteele L, Debels A. Attrition in panel data: the effectiveness of weighting. Eur Sociol Rev. 2007;23(1):81–97.

19. Salon D, Conway MW, Silva DC da, Chauhan RS, Derrible S, Mohammadian A (Kouros), et al. The potential stickiness of pandemic-induced behavior changes in the United States. Proc Natl Acad Sci [Internet]. 2021 Jul 6 [cited 2021 Oct 11];118(27). Available from: https://www.pnas.org/content/118/27/e2106499118

20. Capasso da Silva D, Khoeini S, Salon D, Conway MW, Chauhan RS, Pendyala R, et al. How are Attitudes Toward COVID-19 Associated with Traveler Behavior During the Pandemic? Findings. 2021;

21. Mirtich L, Conway MW, Salon D, Kedron P, Chauhan RS, Derrible S, et al. How Stable Are Transport-Related Attitudes over Time? Findings. 2021 Jun 15;24556.

22. Chauhan RS, Capasso da Silva D, Salon D, Shamshiripour A, Rahimi E, Sutradhar U, et al. COVID-19 related Attitudes and Risk Perceptions across Urban, Rural, and Suburban Areas in the United States. Findings. 2021 06-07;

23. Mohammadi M (Yalda), Rahimi E, Davatgari A, Javadinasr M, Mohammadian A (Kouros), Bhagat-Conway MW, et al. Examining the persistence of telecommuting after the COVID-19 pandemic. Transp Lett [Internet]. 2022 May 26 [cited 2022 Jul 19]; Available from: https://www.tandfonline.com/doi/abs/10.1080/19427867.2022.2077582

24. Javadinasr M, Magassy TB, Rahimi E, Mohammadi M (Yalda), Davatgari A, Mohammadian A (Kouros), et al. Observed and Expected Impacts of COVID-19 on Travel Behavior in the United States: A Panel Study Analysis. In 2022 [cited 2022 Jul 19]. Available from: https://trid.trb.org/view/1926666

25. van der Ploeg T, Austin PC, Steyerberg EW. Modern modelling techniques are data hungry: a simulation study for predicting dichotomous endpoints. BMC Med Res Methodol. 2014;14(1):1–13.

26. Chauhan R, Shi Y, Bartlett A, Sadek AW. Short-Term Traffic Delay Prediction at the Niagara Frontier Border Crossings: Comparing Deep Learning and Statistical Modeling Approaches. J Big Data Anal Transp. 2020 Aug 1;2(2):93–114.

27. Parsa AB, Movahedi A, Taghipour H, Derrible S, Mohammadian AK. Toward safer highways, application of XGBoost and SHAP for real-time accident detection and feature analysis. Accid Anal Prev. 2020;136:105405.

28. Chauhan RS. Short-Term Traffic Delay Prediction at the Niagara Frontier Border Crossings Using Deep Learning. 2019;